\documentstyle[12pt,epsf]{article}
\begin{document}
\newcounter{myfn}[page]
\renewcommand{\thefootnote}{\fnsymbol{footnote}}
\newcommand{\myfootnote}[1]{\setcounter{footnote}{\value{footnote}}%
\footnote{#1}\stepcounter{footnote}}
\renewcommand{\theequation}{\thesection.\arabic{equation}}
\newcounter{saveeqn}
\newcommand{\add}{\addtocounter{equation}{1}}
\newcommand{\alpheqn}{\setcounter{saveeqn}{\value{equation}}%
\setcounter{equation}{0}%
\renewcommand{\theequation}{\mbox{\thesection.\arabic{saveeqn}{\alph{equation}}}}}
\newcommand{\reseteqn}{\setcounter{equation}{\value{saveeqn}}%
\renewcommand{\theequation}{\thesection.\arabic{equation}}}
\newenvironment{nedalph}{\add\alpheqn\begin{eqnarray}}{\end{eqnarray}\reseteqn}
\thispagestyle{empty}
\begin{flushright}
hep-lat/9803026\\
HD--THEP-98-12
\end{flushright}
\vspace{1.5cm}
\begin{center}
{\large\bf{A New Look at the Axial Anomaly in Lattice QED with Wilson Fermions}}\\
\vspace{1.5cm}
Heinz J. Rothe\footnote{\normalsize{Electronic address: h.rothe@thphys.uni-heidelberg.de}} and  N\'eda Sadooghi\footnote{\normalsize{Electronic address: n.sadooghi@thphys.uni-heidelberg.de}}  \\
\vspace{0.5cm}
{\sl{ Institut f\"ur Theoretische Physik}}\\
{\sl{Universit\"at Heidelberg}}\\
{\sl{Philosophenweg 16, 69120 Heidelberg, Germany}}\\
\end{center}
\vspace{1cm}
\begin{center}
{\bf {Abstract}}
\end{center}
\begin{quote}
By carrying out a systematic expansion of Feynman integrals in the lattice spacing, we show that the axial anomaly in the $U\left(1\right)$ lattice gauge theory with Wilson fermions, as determined in one-loop order from an irrelevant lattice operator in the Ward identity, must necessarily be identical to that computed from the dimensionally regulated continuum Feynman integrals for the triangle diagrams.
\end{quote}
\hspace{2cm}
\newpage
\section{Introduction}
\setcounter{footnote}{0}
As is well-known, the Adler-Bell-Jackiw anomaly \cite{adler} in the divergence of the axial vector current in QED, and in the axial flavor singlet current in QCD, is intimately connected with the fact that a shift of  the integration variable in the  linearly divergent Feynman integral for the triangle graph does not leave the integral invariant. An additional  surface term is generated by the shift which, if one requires vector current conservation,  leads to an anomalous axial vector Ward identity. By introducing a Pauli-Villars regularization such shifts in momentum variables are allowed, but chiral symmetry is explicitly broken by the Pauli-Villars regulator masses. Alternatively one can use the dimensional regularization scheme of  't Hooft and Veltman \cite{THOOFT}  to compute the anomalous contribution to the divergence of the axial vector current.
\par
An alternative regularization scheme is that provided by the lattice formulation of QED and QCD. Using Wilson fermions \cite{wil} one avoids the fermion doubling problem at the expense of breaking chiral symmetry explicitly, as demanded by the Nielson-Ninomiya theorem \cite{NN}. The term in the action which is responsible for lifting the fermion degeneracy is a so called irrelevant term which vanishes in the naive continuum limit. Since the lattice regularization is already introduced on the level of the generating functional, the lattice Ward identities for the divergence of the vector and axial vector currents are expected to hold for any finite lattice spacing. While it follows that the divergence of the vector current is conserved, the divergence of the axial vector current is not. The corresponding lattice Ward identity yields an explicit expression for the anomalous contribution, which is entirely determined from the Wilson term in the action that breaks the chiral symmetry explicitly. This anomalous contribution, which is given by an {\it irrelevant} operator, has been calculated a long time ago by Karsten and Smit \cite{KS} and was shown to yield the correct anomaly in the continuum limit.  The axial anomaly on the lattice has been discussed subsequently by various authors \cite{anomal}.  
\par
In the continuum formulation of QED or QCD the anomaly must  necessarily be computed from the divergence of the axial vector current, since the naive Ward identity does not provide us with an explicit expression, as is the case  on the lattice. One is then led to the conclusion that computing the dimensionally regulated triangle graph for the divergence of the axial vector current in the continuum formulation is equivalent to taking the continuum limit of a lattice Feynman integral having no continuum analog. The fact that the anomaly computed from an {\it irrelevant} lattice Feynman integral yields the triangle anomaly of the continuum formulation is not obvious. There is no {\it a priori} reason why this should be the case. It is the purpose of this paper to show why this is {\it necessarily} so. 
\par
The paper is organized as follows. In section 2 we consider the axial vector Ward identity for the $U\left(1\right)$ lattice gauge theory with Wilson fermions, and check explicitly that the anomalous contribution to the divergence of the axial vector current is given by an {\it irrelevant} lattice integral which has precisely the form demanded by the Ward identity. An important ingredient in the proof is that lattice Feynman integrals are invariant under shifts in integration variables. In the continuum formulation, on the other hand, such shifts in linearly divergent Feynman integrals generate an additional surface term which give rise to the anomaly, as we have already remarked before. It is this observation which provides us with the key for proving that the continuum limit of the anomalous contribution to the Ward identity is necessarily given by the triangle graphs in the continuum formulation. The basic idea is that, having proved that the Ward identity is satisfied for any finite lattice spacing, we can also write the {\it irrelevant} anomalous term in the lattice Ward identity as the sum of contributions of the individual Feynman diagrams to the divergence of the axial vector current {\it without} making a shift of integration variables. This sum will now include, besides the additional diagrams arising from irrelevant lattice vertices, the lattice version of the familiar triangle diagrams which do possess a naive continuum limit.  In section 3 we then apply the small-$a$ expansion scheme of Ref. \cite{wet} to this  expression and show that for vanishing lattice spacing it reduces to the dimensionally regulated integrals for the triangle diagrams  of the continuum formulation.  In order to point out a similarity between  the mechanism by which the anomaly is generated on the lattice and in the continuum  formulation, we shall apply the small-$a$ expansion scheme also to the alternative expression given by the {\it irrelevant}  anomalous term in the lattice Ward identity. Section 4  summarizes our results.
\section{Check of the lattice Ward identity}
Consider the action for lattice QED with Wilson fermions
\begin{eqnarray}\label{P21}
\lefteqn{\hspace{-0.5cm}S_{F}^{Wilson}[\overline{\psi},\psi]=\left(m+\frac{4r}{a}\right)\sum\limits_{x}\overline{\psi}\left(x\right)\psi\left(x\right)}\nonumber\\
&&\hspace{-1cm}-\frac{1}{2a}\sum\limits_{x,\mu}\bigg[\overline{\psi}\left(x\right)\left(r-\gamma_{\mu}\right)U_{\mu}\left(x\right)\psi\left(x+a\hat{\mu}\right)+\overline{\psi}\left(x+a\hat{\mu}\right)\left(r+\gamma_{\mu}\right)U_{\mu}^{\dagger}\left(x\right)\psi\left(x\right)
\bigg],
\end{eqnarray}
where $\sum_{x}=\sum_{n}a^{4}$, $m$ is  the fermion mass, $r$ the Wilson parameter and $a$ the lattice spacing. From the invariance of the partition function under the change of variables $\psi\left(x\right)\to \mbox{exp}[i\Lambda\left(x\right)\gamma_{5}]\ \psi\left(x\right)$, and $\overline{\psi}\left(x\right)\to \overline{\psi}\left(x\right)\ \mbox{exp}[i\Lambda\left(x\right)\gamma_{5}]$ one readily derives the following Ward identity for the expectation value of the divergence of the axial vector current in an external field $\{U_{\mu}\left(x\right)\}$,
\begin{nedalph}\label{P22a}
<\Delta_{\mu}^{L}J_{\mu}^{\left(5\right)}\left(x\right)>=2m<\overline{\psi}\left(x\right)\gamma_{5}\psi\left(x\right)>+<X\left(x\right)>,
\end{eqnarray}
where $\Delta_{\mu}^{L}$ is the {\it left} lattice derivative. The axial vector current is given by
\begin{eqnarray}\label{P22b}
J_{\mu}^{\left(5\right)}\left(x\right)=\frac{1}{2}\bigg[\overline{\psi}\left(x\right)\gamma_{\mu}\gamma_{5}U_{\mu}\left(x\right)\psi\left(x+a\hat{\mu}\right)+\overline{\psi}\left(x+a\hat{\mu}\right)\gamma_{\mu}\gamma_{5}U_{\mu}^{\dagger}\left(x\right)\psi\left(x\right)\bigg],
\end{eqnarray}
and the {\it irrelevant} operator $X\left(x\right)$ by
\begin{eqnarray}\label{P22c}
\lefteqn{\hspace{-0.5cm}X\left(x\right)\equiv -\frac{r}{2a}\sum\limits_{\mu}\overline{\psi}\left(x\right)\gamma_{5}\bigg[U_{\mu}\left(x\right)\psi\left(x+a\hat{\mu}\right)+U_{\mu}^{\dagger}\left(x-a\hat{\mu}\right)\psi\left(x-a\hat{\mu}\right)-2\psi\left(x\right)\bigg]
}\nonumber\\
&&\hspace{-1cm}-\frac{r}{2a}\sum\limits_{\mu}\bigg[\overline{\psi}\left(x-a\hat{\mu}\right)U_{\mu}\left(x-a\hat{\mu}\right)+\overline{\psi}\left(x+a\hat{\mu}\right)U_{\mu}^{\dagger}\left(x\right)-2\overline{\psi}\left(x\right)\bigg]\gamma_{5}\psi\left(x\right).
\end{nedalph}\par\noindent
On the other hand, the vector Ward identity reads
\begin{nedalph}\label{P23a}
<\Delta_{\mu}^{L}J_{\mu}\left(x\right)>=0,
\end{eqnarray}
where 
\begin{eqnarray}\label{P23b}
\hspace{-0.4cm}J_{\mu}\left(x\right)=\frac{1}{2}\bigg[\overline{\psi}\left(x+a\hat{\mu}\right)\left(r+\gamma_{\mu}\right)U_{\mu}^{\dagger}\left(x\right)\psi\left(x\right)-\overline{\psi}\left(x\right)\left(r-\gamma_{\mu}\right)U_{\mu}\left(x\right)\psi\left(x+a\hat{\mu}\right)\bigg].
\end{nedalph}\par\noindent
Since the lattice regularization is introduced (non-perturbatively) at the level of the partition function, the Ward identities (\ref{P22a}) and (\ref{P23a}) should hold for any finite lattice spacing. If so, one can calculate the anomalous contribution to the divergence of the axial vector current directly from $<X\left(x\right)>$, as was done originally in Ref. \cite{KS}. This contribution is proportional to the Wilson parameter $r$, with $X\left(x\right)$ an {\it irrelevant} operator that vanishes in the naive continuum limit. The explicit computation of $<X\left(x\right)>$ turns out to yield the correct anomaly
for vanishing lattice spacing \cite{KS}.  
\par
In the continuum formulation, on the other hand, the anomalous Ward identity must necessarily be computed from $<\partial _{\mu}j_{\mu}^{\left(5\right)}\left(x\right)>$ with $j_{\mu}^{\left(5\right)}\left(x\right)=\overline{\psi}\left(x\right)\gamma_{\mu}\gamma_{5}\psi\left(x\right)$. We shall follow here the same procedure, and verify in one-loop order that this external field expectation value is in fact given by the RHS of eq. (\ref{P22a}) for any finite lattice spacing and Wilson parameter $r$.  The purpose of this exercise is to show that the proof, that the anomalous part of $<\Delta_{\mu}^{L}J_{\mu}^{\left(5\right)}\left(x\right)>$ is actually given by an {\it irrelevant} Feynman integral,   
rests heavily on the fact that lattice regulated Feynman integrals are invariant under shifts in integration variables (which is not the case, e.g., for linearly divergent continuum Feynman integrals). It is this observation which will provide us with the key to establish the equality between the anomaly calculated from the dimensionally regulated triangle graph in the continuum formulation, and the anomaly obtained from the expectation value of the {\it irrelevant} operator $X\left(x\right)$ for vanishing lattice spacing. At the same time our check of the lattice Ward identity gives  a nice example of the r\^ole played by irrelevant vertices in ensuring eq. (\ref{P22a}). 
\par
Consider first the LHS of the Ward identity (\ref{P22a}). By setting $U_{\mu}\left(x\right)=e^{ieaA_{\mu}\left(x+a\hat{\mu}/2\right)}$ in eq. (\ref{P22b}), and expanding this expression up to second order in the potentials $A_{\mu}$ (notice that they are defined at the midpoints of the links) one is led in one-loop order to the Feynman diagrams depicted in Fig. 1. The fermion propagator on the lattice reads
\begin{nedalph}\label{P24a}
D^{-1}\left(q\right)=\left(M\left(q\right)+\frac{i}{a}\sum\limits_{\rho}\gamma_{\rho}\sin q_{\rho}a\right)^{-1},
\end{eqnarray}
with $M\left(q\right)$ the momentum dependent Wilson mass,
\begin{eqnarray}\label{P24b}
M\left(q\right)=m+\frac{r}{a}\sum\limits_{\rho}\left(1-\cos {q}_{\rho}a\right).
\end{nedalph}\par\noindent
The lattice expressions for the vertices coupling one or two photons to the vector current are given by
\begin{nedalph}\label{P25a}
V_{\lambda}^{\left(1\right)}\left(\overline{p}\right)=-ie\bigg(\gamma_{\lambda}\cos\frac{\overline{p}_{\lambda}a}{2}-ir\sin\frac{\overline{p}_{\lambda}a}{2}\bigg),
\end{eqnarray}
\begin{eqnarray}\label{P25b}
V_{\lambda\nu}^{\left(2\right)}\left(\overline{p}\right)=ie^{2}a\delta_{\lambda\nu}\left(\gamma_{\lambda}\sin\frac{\overline{p}_{\lambda}a}{2}+ir\cos\frac{\overline{p}_{\lambda}a}{2}\right),
\end{nedalph}\par\noindent
where $\overline{p}=p+p'$, with $p$ ($p'$) the incoming (outgoing) fermion momentum at the vertex. 
In the literature it is customary to write the external field expectation value of $\Delta_{\mu}^{L}J_{\mu}^{\left(5\right)}\left(x\right)$ in the form
\begin{eqnarray}\label{P26}
-\frac{2i}{a}\sum\limits_{\mu}\sin\frac{\left(k+k'\right)_{\mu}a}{2}\Gamma_{\mu\lambda\nu}\left(k,k';a\right)\equiv \tilde{\Gamma}_{\lambda\nu}\left(k,k';a\right),
\end{eqnarray}
where $k$ and $k'$ are the outgoing photon momenta, and  $\Gamma_{\mu\lambda\nu}$ is computed with the following expressions for the axial vector vertices, denoted by dots in Fig. 1,  involving no, one or two photons:
\begin{nedalph}\label{P27a}
v_{5\mu}^{\left(0\right)}\left(\overline{p}\right)=\gamma_{\mu}\gamma_{5}\cos\frac{\overline{p}_{\mu}a}{2},
\end{eqnarray}
\begin{eqnarray}\label{P27b}
v_{5\mu,\lambda}^{\left(1\right)}\left(\overline{p}\right)=-ae\delta_{\mu\lambda}\gamma_{\mu}\gamma_{5}\sin \frac{\overline{p}_{\mu}a}{2},
\end{eqnarray}
\begin{eqnarray}\label{P27c}
v_{5\mu,\lambda\nu}^{\left(2\right)}\left(\overline{p}\right)=-\frac{1}{2}a^{2}e^{2}\delta_{\mu\lambda}\delta_{\mu\nu}\gamma_{\mu}\gamma_ {5}\cos\frac{ \overline{p}_{\mu}a}{2}.
\end{nedalph}\par\noindent
For the following discussion it turns out however to be important to consider the corresponding axial vector vertices for $ \tilde{\Gamma}_{\lambda\nu}$ written in a particular convenient way. These can be obtained directly from the definition of the current (\ref{P22b}): 
\begin{nedalph}\label{P28a}
w^{\left(0\right)}\left(p,p';a\right)=ia^{-1}\sum\limits_{\mu}\gamma_{\mu}\gamma_{5}\left(\sin p_{\mu}a-\sin p'_{\mu}a\right).
\end{eqnarray}
\begin{eqnarray}\label{P28b}
w_{\lambda}^{\left(1\right)}\left(p,p',k;a\right)=ie\gamma_{\lambda}\gamma_{5}\bigg(
\cos\left(2p-k\right)_{\lambda}\frac{a}{2}-\cos\left(2p'+k\right)_{\lambda}\frac{a}{2} \bigg).
\end{eqnarray}
\begin{eqnarray}\label{P28c}
\hspace{-0.2cm}
w_{\lambda\nu}^{\left(2\right)}\left(p,p',k,k';a\right)=-\frac{i}{2}ae^{2}\gamma_{\lambda}\gamma_{5}\delta_{\lambda\nu}\bigg(
\sin\left(2p-k-k'\right)_{\lambda}\frac{a}{2}-\sin\left(2p'+k+k'\right)_{\lambda}\frac{a}{2} 
\bigg)
\end{nedalph}\par\noindent
It can be easily verified, making use of trigonometric relations, that the lattice Feynman integrals corresponding to diagrams (a)-(f) in Fig. 1, computed with these expressions, can be written in the form (\ref{P26}) with the effective  axial vertices  (\ref{P27a})-(\ref{P27c}) for $\Gamma_{\mu\lambda\nu}$. 
\par
Consider now the contribution of diagrams (f) and (c). Clearly diagram (f) does not contribute to the divergence of  $J_{\mu}^{\left(5\right)}\left(x\right)$. Diagram (c) can also be easily shown to vanish by making use of the following identity for (\ref{P28a}) 
\begin{eqnarray}\label{P29}
ia^{-1}\sum\limits_{\mu}\gamma_{\mu}\gamma_{5}\left(\sin p_{\mu}a-\sin p'_{\mu}a\right)=-\gamma_{5}D\left(p\right)-D\left(p'\right)\gamma_{5}+\gamma_{5}\big[M\left(p\right)+M\left(p'\right)\big].
\end{eqnarray}
Next consider the contributions arising from diagrams (a), (b), (d) and (e). Making again use of the identity (\ref{P29}) we can write the contribution of diagrams (a) and (b) in the form
\begin{nedalph}\label{P210a}
T_{\lambda\nu}^{\left(a+b\right)}\left(k,k';a\right)=\int\limits_{-\frac{\pi}{a}}^{+\frac{\pi}{a}}\frac{d^{4}\ell}{\left(2\pi\right)^{4}} N_{\lambda\nu}\left(\ell,k,k';a\right)+{\cal{M}}_{\lambda\nu}\left(k,k';a\right),
\end{eqnarray}
with  
\begin{eqnarray}\label{P210b}
N_{\lambda\nu}\left(\ell,k,k';a\right)&=&\bigg[\mbox{Tr}\bigg(\gamma_{5}V^{\left(1\right)}_{\lambda}\left(2\ell-k\right)D^{-1}\left(\ell\right)V^{\left(1\right)}_{\nu}\left(2\ell+k'\right)D^{-1}\left(\ell+k'\right)\bigg)\nonumber\\
&&+\mbox{Tr}\bigg(\gamma_{5}D^{-1}\left(\ell-k\right)V^{\left(1\right)}_{\lambda}\left(2\ell-k\right)D^{-1}\left(\ell\right)V^{\left(1\right)}_{\nu}\left(2\ell+k'\right)\bigg)\bigg]\nonumber\\
&&+[\left(k,\lambda\right)\leftrightarrow\left(k',\nu\right)],
\end{eqnarray}
and
\begin{eqnarray}\label{P210c}
\lefteqn{
{\cal{M}}_{\lambda\nu}\left(k,k';a\right)=-\int\limits_{-\frac{\pi}{a}}^{+\frac{\pi}{a}}\frac{d^{4}\ell}{\left(2\pi\right)^{4}}\bigg[\bigg(M\left(\ell-k\right)+M\left(\ell+k'\right)\bigg)
}\nonumber\\
&&\hspace{1.5cm}\times\mbox{Tr}\bigg(\gamma_{5}D^{-1}\left(\ell-k\right)V^{\left(1\right)}_{\lambda}\left(2\ell-k\right)
D^{-1}\left(\ell\right)V^{\left(1\right)}_{\nu}\left(2\ell+k'\right)D^{-1}\left(\ell+k'\right)\bigg)\bigg]\nonumber\\
&&\hspace{2.2cm}+[\left(k,\lambda\right)\leftrightarrow\left(k',\nu\right)].
\end{nedalph}\par\noindent
Here $[\left(k,\lambda\right)\leftrightarrow\left(k',\nu\right)]$ denotes the contribution of diagram (b).  Since the term proportional to the Wilson parameter in the Vertex $V^{\left(1\right)}_{\lambda}\left(\overline{p}\right)$ [cf. eq (\ref{P25a})] does not contribute to the trace in $N_{\lambda\nu}\left(\ell,k,k';a\right)$, we can make the following replacement in eq. (\ref{P210b})
\begin{nedalph}\label{P211a}
V_{\lambda}^{\left(1\right)}\left(\overline{p}\right)\longrightarrow -ie\big[\gamma_{\lambda}+\Omega_{\lambda}\left(\overline{p}\right)\big],
\end{eqnarray}
where
\begin{eqnarray}\label{P211b}
\Omega_{\lambda}\left(\overline{p}\right)=-2\gamma_{\lambda}\sin^{2}\frac{\overline{p}_{\lambda}a}{4}
\end{nedalph}\par\noindent
vanishes in the continuum limit.  Hence we obtain 
\begin{eqnarray}\label{P212}
\lefteqn{
N_{\lambda\nu}\left(\ell,k,k';a\right)=}\nonumber\\
&=&\left(-ie\right)^{2}\bigg[\mbox{Tr}\bigg(\gamma_{5}\big[\gamma_{\lambda}+\Omega_{\lambda}\left(2\ell-k\right)\big]D^{-1}\left(\ell\right)\big[\gamma_{\nu}+\Omega_{\nu}\left(2\ell+k'\right)\big]D^{-1}\left(\ell+k'\right)\bigg) 
\nonumber\\
&&\hspace{1.1cm}+\mbox{Tr}\bigg(\gamma_{5}D^{-1}\left(\ell-k\right)\big[\gamma_{\lambda}+\Omega_{\lambda}\left(2\ell-k\right)\big]D^{-1}\left(\ell\right)\big[\gamma_{\nu}+\Omega_{\nu}\left(2\ell+k'\right)\big]\bigg)
\bigg]\nonumber\\
&&+[\left(k,\lambda\right)\leftrightarrow\left(k',\nu\right)]. 
\end{eqnarray}
This expression can be decomposed into irrelevant terms (i.e., vanishing in the continuum limit) proportional to one or two $\Omega_{\mu}\left(q\right)$'s, and terms which possess a continuum analog. By making appropriate shifts of integration variables these latter terms can be readily shown not to contribute to the integral (\ref{P210a}). In the continuum formulation the very same shifts would also lead naively to a vanishing result, but because one is faced with linearly divergent integrals a surface term is generated which is nothing else but the anomaly. In a lattice regulated theory such shifts of variables leave the integrals invariant and the anomaly is actually generated only by ${\cal{M}}_{\lambda\nu}\left(k,k';a\right)$, given  in eq. (\ref{P210c}). In fact, as we now show, the contributions of the irrelevant terms in eq. (\ref{P212}) to the integral (\ref{P210a}) are canceled by  the contributions arising from the lattice diagrams (d) and (e) involving the irrelevant vertex $w_{\lambda}^{\left(1\right)}\left(p,p',k;a\right)$ defined in eq. (\ref{P28b}). Writing $w_{\lambda}^{\left(1\right)}\left(p,p',k;a\right)$ in the form
\begin{eqnarray}\label{P213}
w_{\lambda}^{\left(1\right)}\left(p,p',k;a\right)=-ie\gamma_{5}\bigg[\Omega_{\lambda}\left(2p-k\right)-\Omega_{\lambda}\left(2p'+k\right)\bigg],
\end{eqnarray}
where $\Omega_{\lambda}\left(q\right)$ has been defined in eq. (\ref{P211b}), and noting that we can make again the replacement (\ref{P211a}) for the vertex coupling the photon to the vector current (the term proportional to $r$ in eq. (\ref{P25a})  does not contribute to the trace), we obtain the following expression for the contributions of diagrams (d) and (e)
\begin{eqnarray}\label{P214}
\lefteqn{\hspace{-1cm}T_{\lambda\nu}^{\left(d+e\right)}\left(k,k';a\right)=e^{2}\int\limits_{-\frac{\pi}{a}}^{+\frac{\pi}{a}}
\bigg[\mbox{Tr}\Bigg(\gamma_{5}\big[\Omega_{\nu}\left(2\ell-k'\right)-\Omega_{\nu}\left(2\ell+2k+k'\right)\big]D^{-1}\left(\ell\right)}\nonumber\\
&&\hspace{2cm}\times\big[\gamma_{\lambda}+\Omega_{\lambda}\left(2\ell+k\right)\big]D^{-1}\left(\ell+k\right)\Bigg)\bigg]+[\left(k,\lambda\right)\leftrightarrow\left(k',\nu\right)].
\end{eqnarray} 
By appropriate shifts of integration variables one then readily verifies that the contribution of these diagrams cancel the remaining irrelevant contributions in eq. (\ref{P212}) to  the integral in (\ref{P210a}). Hence the LHS of the Ward identity (\ref{P22a}) is given just by ${\cal{M}}_{\lambda\nu}$, defined in eq. (\ref{P210c}), i.e. by an {\it irrelevant} lattice integral. It is now easy to check that this result for $<\Delta_{\mu}^{L}J_{\mu}^{\left(5\right)}\left(x\right)>$ coincides with that computed from the RHS of the Ward identity (\ref{P22a}). Since the vertices associated with the {\it irrelevant} operator $X\left(x\right)$, defined in eq. (\ref{P22c}), are proportional to $\gamma_{5}$, only diagrams (a) and (b) will now contribute because of the symmetry properties of  the Dirac trace. The corresponding $\gamma_{5}$-vertex is given by
\begin{nedalph}\label{P215a}
v\left(p,p';a\right)=\gamma_{5}\bigg[{\cal{M}}_{r}\left(p\right)+{\cal{M}}_{r}\left(p'\right)\bigg],
\end{eqnarray}    
where 
\begin{eqnarray}\label{P215b}
{\cal{M}}_{r}\left(q\right)=\frac{r}{a}\sum\limits_{\rho}\left(1-\cos q_{\rho}a\right)
\end{nedalph}\par\noindent
is the momentum dependent Wilson mass (\ref{P24b}) evaluated for $m=0$. Including the contribution of $2m<\overline{\psi}\left(x\right)\gamma_{5}\psi\left(x\right)>$, the Feynman integral corresponding to the RHS of eq. (\ref{P22a}) is therefore given by (\ref{P210c}).  
\par
Notice that only after using the identity (\ref{P29}), and making appropriate shifts in integration variables, we were led to the conclusion that the contribution of diagrams (a)-(f) to the anomalous part of $<\Delta_{\mu}^{L}J_{\mu}^{\left(5\right)}\left(x\right)>$ combined to yield the expectation value of the {\it irrelevant} operator  $X\left(x\right)$. On the other hand, the divergence of the axial vector current written in the form (\ref{P26}) with the effective axial vertices (\ref{P27a})-(\ref{P27c}) for $\Gamma_{\mu\lambda\nu}$, does not only include the irrelevant contributions of the diagrams (c)-(f), but also the contributions of diagrams (a) and (b) which do possess a continuum analog. Since we have shown the Ward identity to be satisfied for any finite lattice spacing, we have therefore  the following relation 
\begin{eqnarray}\label{P216}
\tilde{\Gamma}_{\lambda\nu}\left(k,k';a\right)={\cal{M}}_{\lambda\nu}\left(k,k';a\right),
\end{eqnarray}
where $\tilde{\Gamma}_{\lambda\nu}\left(k,k';a\right)$ and ${\cal{M}}_{\lambda\nu}\left(k,k';a\right)$ have been defined in eqs. (\ref{P26}) and (\ref{P210c}), respectively. In the following section we will now use the small-$a$ expansion scheme of Ref. \cite{wet} to show that in the continuum limit $\tilde{\Gamma}_{\lambda\nu}$ reduces to the dimensionally regulated Feynman integral for the triangle  diagrams of the continuum formulation.   
\section{Computation of the axial anomaly in the small-$a$ expansion scheme}
\setcounter{equation}{0}
In the small-$a$ expansion scheme of Ref. \cite{wet} one departs from a $D$-dimensionally regulated one-loop lattice Feynman integral having the generic form
\begin{eqnarray}\label{P31}
F\left(p,m;a,D\right)=\int\limits_{-\frac{\pi}{a}}^{+\frac{\pi}{a}}\frac{d^{D}\ell}{\left(2\pi\right)^{D}}\ H\left(\ell,p,m;a,D\right),
\end{eqnarray}
where $p$ and $m$ stand collectively for the set of external momenta and masses, and it is understood that the coupling constant $e$ is to be replaced, as is  usual in the dimensional regularization scheme, by $e\mu^{\left(4-D\right)/2}$, with $\mu$ an arbitrary mass scale.  An important feature of the lattice Feynman integral is that the lattice spacing appears in the integrand as well as in the integration limits. This integrand is a periodic function of $p$ and $p'$ and reduces to the integrand of the corresponding continuum Feynman integral in the limit of vanishing lattice spacing.
\par
According to the small-$a$ expansion scheme $F\left(p,m;a,D\right)$ has the following power series expansion in the lattice spacing:\footnote{The small-$a$ expansion has been applied also in \cite{wet, ned}.}
\begin{eqnarray}\label{P32}
\lefteqn{F\left(p,m;a,D=4\right)=\lim\limits_{D\to 4}\bigg[
T_{J}\int\limits_{-\infty}^{+\infty}\frac{d^{D}\ell}{\left(2\pi\right)^{D}}\ H\left(\ell,p,m;a,D\right)}\nonumber\\
&& \hspace{-0.5cm}+a^{d_{H}-D}T_{J-\left(d_{H}-4\right)}\int\limits_{-\pi}^{+\pi}\frac{d^{D}\hat{\ell}}{\left(2\pi\right)^{D}}\ H\left(\hat{\ell},ap,am;a=1,D\right)
\bigg]+O\left(a^{J+\left(4-D\right)+1}\right),
\end{eqnarray}
where $d_{H}$ is the inverse-mass dimension of $H\left(\ell,p,m;a,D\right)$, $\hat{\ell}$ is the dimensionless loop variable, and $T_{k}f\left(a\right) $ denotes the Taylor  expansion of $f\left(a\right)$ around  $a=0$ up to $O\left(a^{k}\right)$. Note that the integration in the first integral extends, as in the continuum, over an infinite range, and that the leading $O\left(a^{0}\right)$ contribution is just the dimensionally regulated continuum Feynman integral. The Taylor expansion of the second integral yields polynomials in the momenta with coefficients having the form of lattice Feynman integrals with periodic integrands. 
\par
In the following we will first use the small-$a$ expansion scheme to calculate the anomaly from eq. (\ref{P210c}).  Although this computation is not required for the main  objective of the paper, it serves to illustrate the distinct ways in which the {\it relevant} operator $2m\overline{\psi}\left(x\right)\gamma_{5}\psi\left(x\right)$, and the {\it irrelevant} operator $X\left(x\right)$ in the Ward identity (\ref{P22a}) are handled by the small-$a$ expansion scheme. At the same time it  will also serve to point out the similarity between the mechanism by which the anomaly is generated on the lattice and in the continuum formulation. 
\vspace{0.3cm}
\par\noindent
{\it i) The anomaly calculated from the small-$a$ expansion of ${\cal{M}}_{\lambda\nu}\left(k,k';a\right)$}
\vspace{0.3cm}
\par
Consider the RHS of the Ward identity (\ref{P22a}). As we have seen, it is given in one-loop order by ${\cal{M}}_{\lambda\nu}\left(k,k';a\right)$, defined in eq.  (\ref{P210c}). Extracting from this expression the contribution proportional to $m$, making use of eqs. (\ref{P24b}) and (\ref{P215b}), we have that
\begin{nedalph}\label{P33a}
\hspace{-0.2cm}
{\cal{M}}_{\lambda\nu}\left(k,k';a\right)=2m\int\limits_{-\frac{\pi}{a}}^{+\frac{\pi}{a}}\frac{d^{D}\ell}{\left(2\pi\right)^{D}} \bigg[I^{\left(a\right)}_{\lambda\nu}\left(\ell,k,k';a\right)+I_{\lambda\nu}^{\left(b\right)}\left(\ell,k,k';a\right) \bigg]+{\cal{A}}_{\lambda\nu}\left(k,k';a\right),
\end{eqnarray}
where
\begin{eqnarray}\label{P33b}
\lefteqn{\hspace{-1cm}
{\cal{A}}_{\lambda\nu}\left(k,k';a\right)=}\nonumber\\
&&\hspace{-1.3cm}=\int\limits_{\frac{\pi}{a}}^{+\frac{\pi}{a}}\frac{d^{D}\ell}{\left(2\pi\right)^{D}}\bigg[\bigg({\cal{M}}_{r}\left(\ell-k\right)+{\cal{M}}_{r}\left(\ell+k'\right)\bigg)I^{\left(a\right)}_{\lambda\nu}\left(\ell,k,k';a\right)\bigg]+[\left(k,\lambda\right)\leftrightarrow\left(k',\nu\right)],
\end{eqnarray}
is the anomalous contribution to the divergence of the axial vector current. Here  
\begin{eqnarray*}
\lefteqn{\hspace{-0.5cm}
I_{\lambda\nu}^{\left(a\right)}\left(\ell,k,k';a\right)=}\nonumber\\
&=&-\mbox{Tr}\bigg(\gamma_{5}D^{-1}\left(\ell-k\right)V^{\left(1\right)}_{\lambda}\left(2\ell-k\right)D^{-1}\left(\ell\right)V_{\nu}^{\left(1\right)}\left(2\ell+k'\right)D^{-1}\left(\ell+k'\right)
\bigg),
\end{eqnarray*}
and 
\begin{eqnarray}\label{P33c}
I_{\lambda\nu}^{\left(a\right)}\left(\ell,k,k';a\right)=I_{\nu\lambda}^{\left(b\right)}\left(\ell,k',k;a\right).
\end{nedalph}\par\noindent
The fermion propagator $D^{-1}\left(q\right)$ and the fermion-photon vertex $V_{\lambda}^{\left(1\right)}\left(\overline{p}\right)$ are defined in eqs.  (\ref{P24a}) and (\ref{P25a}),  respectively. 
We now apply the small-$a$ expansion scheme (\ref{P32}) to this expression,  with $d_{H}=2$ in  the present case. Since it turns out that the corresponding Taylor coefficients in (\ref{P32}), up to   $O\left(a^{0}\right)$, are finite for $D\to 4$, we can set $D=4$ in this equation.
\par
Consider the integral in (\ref{P33a}) whose leading contribution to the first, as well as to the second term in (\ref{P32}) is found to be of $O\left(a^{0}\right)$. The contribution in this order to the second term  vanishes due to the symmetry properties of the Dirac trace. We are therefore only left  in $O\left(a^{0}\right)$ with the contribution of the first term in eq. (\ref{P32}), which   just yields the usual expression for the divergence of the axial vector current arising from the explicit breaking of chiral symmetry due to the fermion mass $m$,\footnote{Although the authors in Ref. \cite{KS} claim, without giving a proof, that they can also generate this contribution, we do not see how this is possible in their approach. In particular we question the remark following eq. (5.42) in their paper.}
\begin{nedalph}\label{P34a}
\lim\limits_{a\to 0}2m\int\limits_{-\frac{\pi}{a}}^{+\frac{\pi}{a}}\frac{d^{4}\ell}{\left(2\pi\right)^{4}}\bigg[I_{\lambda\nu}^{\left(a\right)}\left(\ell,k,k';a\right)+I_{\lambda\nu}^{\left(b\right)}\left(\ell,k,k';a\right)\bigg] =
2mT_{\lambda\nu}\left(k,k'\right),
\end{eqnarray}
where
\begin{eqnarray}\label{P34b}
T_{\lambda\nu}\left(k,k'\right)=\frac{me^{2}}{2\pi^{2}}\varepsilon_{\lambda\nu\mu\rho}k_{\mu}k'_{\rho}\int\limits_{0}^{1}dx\int\limits_{0}^{1-x}dy\ \frac{1}{m^{2}-\left(xk-yk'\right)^{2}+xk^{2}+yk'^{2}}.
\end{nedalph}
\par  
What concerns the anomalous term ${\cal{A}}_{\lambda\nu}\left(k,k';a\right)$,  in eq. (\ref{P33a}), exactly the opposite is true. Since the integrand in (\ref{P33b}) vanishes in the continuum limit, it does not contribute in $O\left(a^{0}\right)$ to the first integral in eq. (\ref{P32}). It does however contribute to the second term in this equation. Although the leading term in the expansion is of $O\left(a^{-2}\right)$ one finds,  after a straightforward, but lengthy calculation, that the coefficients of $a^{-2}$ and $a^{-1}$ vanish, and that
\begin{nedalph}\label{P35a}  
{\cal{A}}_{\lambda\nu}\left(k,k'\right)\equiv\lim\limits_{a\to 0}{\cal{A}}\left(k,k';a\right)=16e^{2}\varepsilon_{\lambda\nu\mu\rho}k_{\mu}k'_{\rho}\int\limits_{-\pi}^{+\pi}\frac{d^{4}\hat{\ell}}{\left(2\pi\right)^{4}} H_{\lambda\nu}\left(\hat{\ell}\right),
\end{eqnarray}
where
\begin{eqnarray}\label{P35b}
H_{\lambda\nu}\left(\hat{\ell}\right)=\frac{\cos\hat{\ell}_{\lambda}\cos\hat{\ell}_{\nu}\cos\hat{\ell}_{\mu}\Bigg(\hat{\cal{M}}_{r}^{2}\left(\hat{\ell}\right) \cos\hat{\ell}_{\rho}-4r\hat{\cal{M}}_{r}\left(\hat{\ell}\right)\sin^{2}\hat{\ell}_{\rho}
\Bigg)}{\bigg[\hat{\cal{M}}_{r}^{2}\left(\hat{\ell}\right)+\sum\limits_{\kappa}\sin^{2}\hat{\ell}_{\kappa}\bigg] ^{3} },
\end{nedalph}\par\noindent
and where all quantities with ''hat'' are measured in lattice units. This coincides with the expression of Karsten and Smit \cite{KS}. After making use of the identity \cite{KS}
\begin{eqnarray}\label{P36}
\lefteqn{\hspace{-1cm}
\hat{\cal{M}}_{r}^{2}\left(\hat{\ell}\right)\cos\hat{\ell}_{\rho}-4r\hat{\cal{M}}_{r}\left(\hat{\ell}\right)\sin^{2}\hat{\ell}_{\rho}
=\cos\hat{\ell}_{\rho}\left(\hat{\cal{M}}_{r}^{2}\left(\hat{\ell}\right)+4\sin^{2}\hat{\ell}_{\rho}\right)}\nonumber\\
&&
+\bigg(\hat{\cal{M}}_{r}^{2}\left(\hat{\ell}\right)+\sum\limits_{\kappa}\sin^{2}\hat{\ell}_{\kappa}\bigg) ^{3}\sin\hat{\ell}_{\rho}\frac{\partial}{\partial \hat{\ell}_{\rho}}\left(\hat{\cal{M}}_{r}^{2}\left(\hat{\ell}\right)+\sum\limits_{\kappa}\sin^{2}\hat{\ell}_{\kappa}\right) ^{-2}
\end{eqnarray} 
and performing a partial integration one readily verifies that 
\begin{nedalph}\label{P37a}
{\cal{A}}_{\lambda\nu}\left(k,k'\right)=\lim\limits_{\epsilon\to 0}\int\limits_{V-V_{\epsilon}}\frac{d^{4}\hat{\ell}}{\left(2\pi\right)^{4}} \hat{\partial}_{\rho}{\cal{K}}_{\rho\lambda\nu}\left(\hat{\ell},k,k'\right)
\end{eqnarray}
where
\begin{eqnarray}\label{P37b}
{\cal{K}}_{\rho\lambda\nu}\left(\hat{\ell},k,k'\right)=16e^{2}\varepsilon_{\lambda\nu\mu\rho}k_{\mu}k'_{\rho}\cos\hat{\ell}_{\lambda}\cos\hat{\ell}_{\nu}\cos\hat{\ell}_{\mu} \frac{\sin\hat{\ell}_{\rho}}{\bigg[\hat{\cal{M}}_{r}^{2}\left(\hat{\ell}\right)+\sum\limits_{\kappa}\sin^{2}\hat{\ell}_{\kappa}\bigg] ^{2}},
\end{nedalph}\par\noindent
 where the integration extends over  a 4-dimensional cube with edges lying in the first Brillouine zone $\big[-\pi,+\pi\big]$, excluding an infinitesimal sphere with radius $\epsilon$  around the infrared singularity at the origin.\footnote{Since the integral in (\ref{P35a}) is infrared finite, we are free to calculate it as the limit $\epsilon\to 0$ of the volume  integral $V-V_{\epsilon}$. This provides a natural regularization of the, in the limit $\epsilon\to 0$, logarithmically divergent integrals appearing after making use of eq. (\ref{P36}).}  By applying Gauss theorem and taking the limit $\epsilon\to 0$, 
one obtains the result of Ref. \cite{KS}
\begin{eqnarray}\label{P38}
{\cal{A}}_{\lambda\nu}\left(k,k'\right)=-\frac{e^{2}}{2\pi^{2}}\varepsilon_{\lambda\nu\mu\rho}k_{\mu}k'_{\rho},
\end{eqnarray}
which is just the well-known anomaly of the continuum formulation. This, as we have already mentioned before, is a remarkable result, since the anomaly has been computed from an expression having no continuum analog. On the other hand, the sum of the two triangle diagrams contributing to the anomaly in continuum QED are known to give a non-vanishing contribution only because linearly divergent Feynman integrals are not invariant under shifts in integration variables. The surface term generated by such shifts, arising from an ultraviolet divergence, yields the anomaly (see  e.g. Ref. \cite{jackiw}).  
On the lattice the surface term is generated by a similar mechanism, but the anomaly arises from an infrared divergence. This, and the above explicit computation, suggests that one can actually prove that the anomaly arising from the explicit breaking of chiral symmetry by the {\it irrelevant}  Wilson term in the action is {\it necessarily} identical with that calculated from the continuum triangle Feynman diagrams.  The crucial ingredient going into the proof is, that since we have shown that the Ward identity (\ref{P22a})  is satisfied for any finite lattice spacing, we can just as well compute the anomaly from the divergence of the axial vector current, using the small-$a$ expansion scheme without making use of the identity (\ref{P29}) and of the invariance of lattice Feynman integrals under shifts in integration variables, i.e. from eq. (\ref{P216}), where $\tilde{\Gamma}_{\lambda\nu}$ has been defined in eq. (\ref{P26}).   Since the Taylor expansion of the individual lattice Feynman integrals contributing to 
$\tilde{\Gamma}_{\lambda\nu}$ are sensitive to shifts in integration variables, we will be able to transfer the anomaly arising from an {\it irrelevant} lattice operator, to the familiar triangle anomaly in the continuum. We now show how the small-$a$ expansion scheme is able to realize this program. 
\vspace{0.3cm}
\par\noindent
{\it ii)  Alternative computation of the anomaly}
\vspace{0.3cm}
\par
Consider the individual Feynman integrals associated with the diagrams (a)-(f) contributing to $\tilde{\Gamma}_{\lambda\nu}$ in eq. (\ref{P26}) with the momenta chosen in the way indicated in the Fig. 1. Since the integrands of the integrals associated with diagrams (a) and (b) possess a continuum limit, while those associated with the remaining diagrams vanish in the limit $a\to 0$ (due to the irrelevant vertices), we immediately conclude that the leading contribution of $O\left(a^{0}\right)$ of the first term in (\ref{P32}) just yields the usual dimensionally regularized continuum expression for the triangle graphs (a) and (b). The computation of the contributions to the second term in this equation, which involves only lattice integrals with periodic integrands, is very tedious. The coefficients of $O\left(a^{-2}\right)$ and $O\left(a^{-1}\right)$ turn out to be finite for $D\to 4$, and vanish because of the symmetry properties of the trace. In $O\left(a^{0}\right)$, however,  one is also faced  with infrared divergent lattice integrals for $D\to 4$  at  intermediate stages of the calculation. Hence one must consider the dimensionally regulated expressions, as demanded by the small-$a$ expansion scheme (\ref{P32}). In the 't Hooft-Veltman scheme \cite{THOOFT} $\{\gamma_{\mu},\gamma_{\nu}\}=2\delta_{\mu\nu}$ remains valid in $D$ dimensions, while $\gamma_{5}$ anticommutes with $\gamma_{\mu}$ for $\mu=1,\cdots,4$, and commutes with $\gamma_{\mu}$ for $\mu>4$. In the case  where one is dealing with convergent lattice integrals for $D\to 4$, one can work directly in four dimensions using the usual anticommutation rules for the $\gamma_{\mu}$- and $\gamma_{5}$-matrices. What concerns the in the limit $D\to 4$ divergent integrals, we found,  after a lengthy computation, involving many partial integrations, and making {\it only} use of the anticommutation relations $\{\gamma_{\mu},\gamma_{\nu}\}=2\delta_{\mu\nu}$, that one is left with  the following expression for the Taylor coefficient for the contribution to the second term in eq. (\ref{P32}) in $O\left(a^{0}\right)$,
\begin{nedalph}\label{P39a}
e^{2}\sum\limits_{\mu}\left(k+k'\right)_{\mu}\bigg[\frac{2}{3}\left(k-k'\right)_{\lambda}\mbox{Tr}\left(\gamma_{5}\gamma_{\nu}\gamma_{\mu}\right) {\cal{P}}_{\mu\lambda\nu}-2im\delta_{\lambda\nu}\mbox{Tr}\left(\gamma_{\mu}\gamma_{5}\right){\cal{Q}}_{\mu\lambda\nu}\bigg],
\end{eqnarray}
where
\begin{eqnarray}\label{P39b}
{\cal{P}}_{\mu\lambda\nu}=\int\limits_{-\pi}^{+\pi}\frac{d^{D}\hat{\ell}}{\left(2\pi\right)^{D}} 
\frac{\cos\hat{\ell}_{\mu}\cos\hat{\ell}_{\lambda}\cos^{2}\hat{\ell}_{\nu}\bigg[-3\hat{\cal{M}}_{r}^{2}\left(\hat{\ell}\right)-4\sin^{2}\hat{\ell}_{\lambda} +\sum\limits_{\rho}\sin^{2}\hat{\ell}_{\rho} \bigg]
}{
\bigg[\hat
{\cal{M}}_{r}^{2}\left(\hat{\ell}\right)+\sum\limits_{\rho}\sin^{2}\hat{\ell}_{\rho}
\bigg]^{3}},
\end{eqnarray}
and
\begin{eqnarray}\label{P39c}
{\cal{Q}}_{\mu\lambda\nu}=\int\limits_{-\pi}^{+\pi}\frac{d^{D}\hat{\ell}}{\left(2\pi\right)^{D}} 
 \frac{\cos\hat{\ell}_{\mu}\cos\hat{\ell}_{\lambda}\cos\hat{\ell}_{\nu}\bigg[4\sin^{2}\hat{\ell}_{\lambda}-\sum\limits_{\rho}\sin^{2}\hat{\ell}_{\rho}\bigg]
}{
\bigg[\hat
{\cal{M}}_{r}^{2}\left(\hat{\ell}\right)+\sum\limits_{\rho}\sin^{2}\hat{\ell}_{\rho}
\bigg]^{3}}.
\end{nedalph}\par\noindent
 Since the traces in eq. (\ref{P39a}) vanish for $D\to 4$, the only non-vanishing contribution to this  coefficient  can arise from the divergent parts of the integrals (\ref{P39b}) and (\ref{P39c}).
Consider first the expression (\ref{P39b}). The potentially divergent contribution to this integral arises from the integration region $\{\hat{\ell}_{\rho}\}\approx 0$, where $\cos \hat{\ell}_{\kappa}\approx 1$. What concerns $\sin^{2}\hat{\ell}_{\kappa}$, it is convenient to write it in the form $\sin^{2}\hat{\ell}_{\kappa}\approx 4\sin^{2}\frac{\hat{\ell}_{\kappa}}{2}$. With these approximations the integral in eq. (\ref{P39b}) becomes independent of $\lambda$, so that we can further replace $\sin^{2}\frac{\hat{\ell}_{\lambda}}{2}$ by $\frac{1}{D}\sum\limits_{\kappa}\sin^{2}\frac{\hat{\ell}_{\kappa}}{2}$. Then the potentially divergent contribution reduces to 
\begin{eqnarray*} 
\frac{D-4}{D}\int\limits_{-\pi}^{+\pi}\frac{d^{D}\hat{\ell}}{\left(2\pi\right)^{D}} \frac{1}{\bigg(4\sum\limits_{\rho}\sin^{2}\frac{\hat{\ell}_{\rho}}{2}\bigg)^{2}}.
\end{eqnarray*}
For $D\to 4$ the divergent part of this integral behaves like $[8\pi^{2}\left(D-4\right)]^{-1}$ (cf. Ref. \cite{ned}). We therefore find that the integral  (\ref{P39b}) is actually convergent for $D\to 4$, so that the first term in (\ref{P39a}) can be evaluated directly in four dimensions and vanishes since $\mbox{Tr}\left(\gamma_{5}\gamma_{\nu}\gamma_{\mu}\right)=0$ for $D=4$. Proceeding in a similar way one finds that the contribution of the second term in eq. (\ref{P39a}) also vanishes for $D\to 4$. 
Hence, in this alternative computation of the anomaly, the second term in eq. (\ref{P32}) does not contribute in the continuum limit, and the anomaly, as well as the contribution (3.4) arising from explicit breaking of the chiral symmetry by the fermion mass, are obtained entirely from the first integral in eq. (\ref{P32}).
\par
We have therefore shown that taking the continuum limit of the anomalous contribution to the divergence of the axial vector current, arising from  the {\it irrelevant} operator $X\left(x\right)$ in the Ward identity (\ref{P22a}), is necessarily equivalent to taking the $D\to 4$ limit of the dimensionally regulated continuum Feynman integrals for the triangle diagrams.
Hence the small-$a$ expansion scheme has allowed us to establish a direct connection between the anomaly calculated from an {\it irrelevant} lattice Feynman integral and  a dimensionally regulated continuum integral. 
\section{Conclusion}
In this paper we have taken a new look at the axial anomaly in lattice QED with Wilson fermions, originally discussed in Ref. \cite{KS}. We have first checked explicitly that the axial Ward identity is satisfied in one-loop order for  any finite lattice spacing and Wilson parameter. In checking the lattice Ward identity, we have made extensive use of the  fact that lattice Feynman integrals are invariant under a shift of integration variables, which is not the case in general for ultraviolet divergent integrals in the continuum formulation. By making use of this  property we have shown explicitly that the anomalous contribution to the divergence of the axial vector current is given in one-loop order by an {\it irrelevant}  lattice Feynman integral.  At the same time  this check provided  a nice example of the r\^ole played by the irrelevant vertices in ensuring the validity of the Ward identity. Having shown that it holds for any finite lattice spacing we were then led to consider an alternative expression for the anomalous part of the divergence of the axial vector current.  By applying  the small-$a$ expansion scheme of Ref. \cite{wet} to this expression  we have then shown in a systematic way that the axial anomaly, as computed from an {\it irrelevant} lattice operator, {\it necessarily} coincides with that computed from the dimensionally regulated triangle diagram in the continuum formulation. The crucial ingredient going into the proof was that the small-$a$ expansion of lattice Feynman integrals is sensitive to shifts in integration variables. This made it possible to transfer the anomaly arising from an {\it irrelevant} lattice operator to the triangle anomaly computed in the continuum formulation of QED in the  dimensional regularization scheme.

\section*{Figure caption}
Figure 1: Lattice diagrams contributing to the divergence of the axial vector current.
\newpage
\epsfxsize12cm                
\begin{figure}[h]          
                             
\leavevmode                 
\centering                      
\epsffile{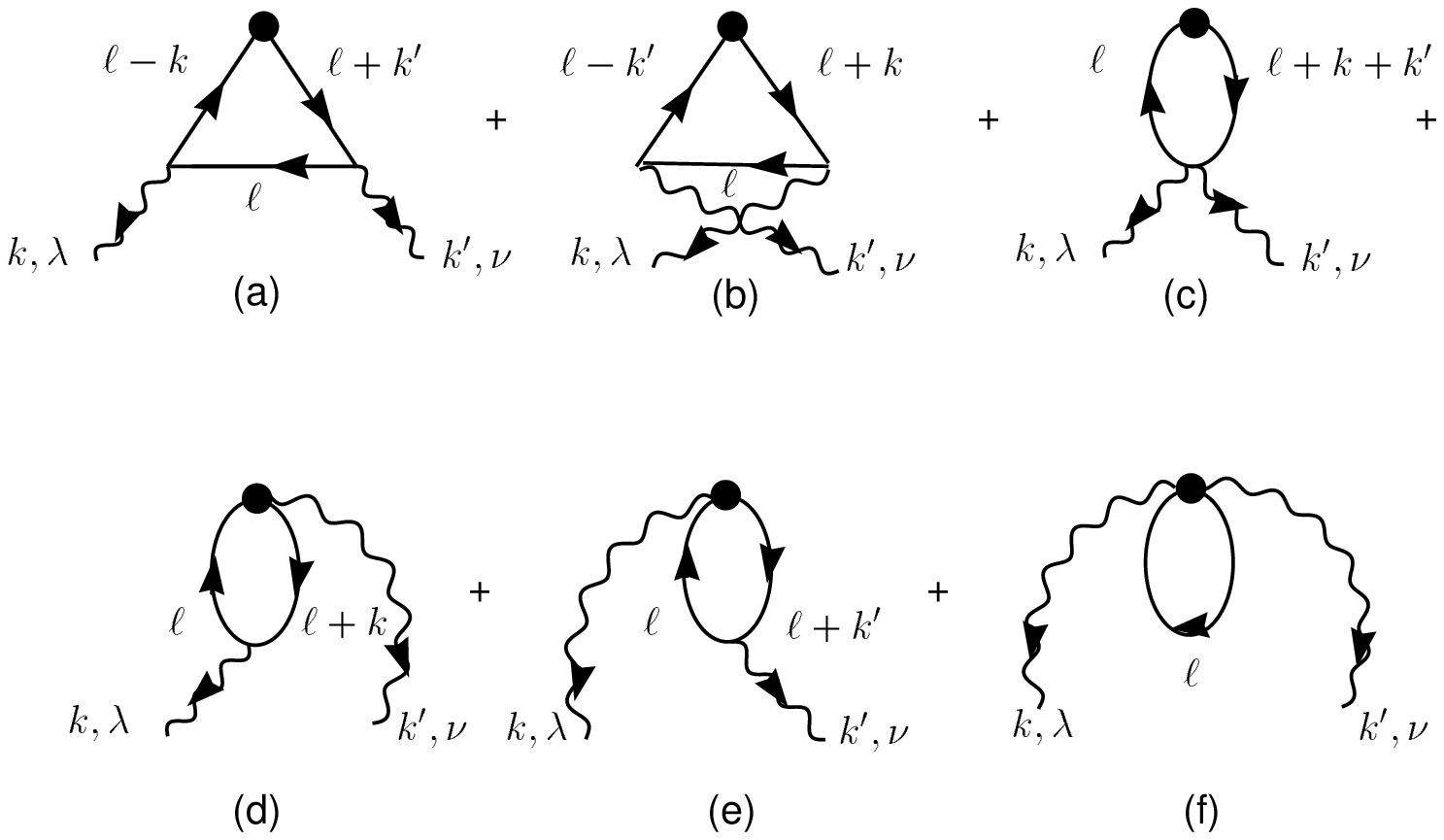}   
\caption{\sl {Lattice diagrams contributing to the divergence of the axial vector current.}}
               
\end{figure}
\parindent0em

\end{document}